\documentclass[cameraready]{Interspeech}

\title{Silence is Golden: Mitigating Hallucinations in Large Audio-Language Models via Layer-Weighted Vector Steering}
\usepackage{multirow}
\usepackage{tikz}
\usetikzlibrary{positioning, arrows.meta, shapes.geometric, calc, shadows, backgrounds, decorations.pathreplacing}
\author[affiliation={1,2}]{Tsung-En}{Lin}
\author[affiliation={1,2}]{Kuan-Yi}{Lee}
\author[affiliation={1,3}]{Hung-Yi}{Lee}

\address{
    $^1$ National Taiwan University, Taipei, Taiwan \\
    $^2$ Internship at ASUS Open Cloud Infrastructure Software Center, Taipei, Taiwan \\
    $^3$ NTU Artificial Intelligence Center of Research Excellence (NTU AI-CoRE), Taipei, Taiwan
}

\email{\{b11901154, b10901091, hungyilee\}@ntu.edu.tw}


\usepackage{comment}

\begin{document}

\maketitle

\begin{abstract}
Large Audio-Language Models (LALMs) excel in Audio QA but often suffer from hallucinations ungrounded in the audio. To our knowledge, we are the first to propose applying vector steering to the audio domain to mitigate this. Unlike text-based steering, our silence-anchored contrastive approach steers the model away from hallucinations by contrasting active audio against a silent baseline. Probing internal states reveals a strong correlation between specific layer representations and output correctness. Leveraging this, we introduce Layer-Weighted Vector Steering (LWVS), a training-free intervention that increases steering strength at influential layers. On the Audio Hallucination QA dataset, LWVS significantly outperforms baselines, boosting Recall on the Gemma model by 15.6\% (53.4\% to 69.0\%). Crucially, MMAU benchmark tests confirm LWVS preserves and even enhances general audio understanding, achieving an 8\% relative accuracy increase on the Qwen model (54.8\% to 59.2\%).
\end{abstract}
\section{Introduction}
Large Audio-Language Models \cite{chu2024qwen2, goel2025audio, team2025gemma, tseng2025taste, wu2025step, deshmukh2023pengi, tang2024salmonn} have extended the capabilities of LLMs to multimodal tasks like audio captioning and question answering. Concurrently, specialized Speech Language Models (SLMs) \cite{lakhotia2021generative, borsos2023audiolm, wang2023neural, zhang2023speechgpt, defossez2024moshi, fang2024llama} have significantly advanced discrete speech modeling and spoken dialogue. However, both architectures frequently suffer from hallucination, generating content ungrounded in the audio input \cite{kuan2024understanding}.

To address this, we draw inspiration from VISTA \cite{vis}, which utilizes activation steering to guide vision language model behavior. However, applying steering vectors derived solely from text or methods that lack audio context is suboptimal for LALMs. To effectively ground the model, we introduce a \textbf{modality-aware} strategy. We construct a steering vector that explicitly contrasts the active audio representation against a \textit{silent audio clip of the same length}. This forces the model to attend to the presence of acoustic information, reducing the likelihood of generating events that do not exist.

To determine the most effective intervention layers, we investigate the internal mechanics of hallucination, drawing on established practices of layer-wise analysis in speech models \cite{pasad2021layer, chen2022wavlm, chung2019unsupervised}. We probe the model's hidden states by calculating the cosine similarity between the original hidden state and our steering vector for each layer. We then compute Cohen's $d$ effect size on these similarity scores to quantify their relationship with output correctness. Our analysis, visualized in Figure \ref{fig:Gemma-layers}, reveals that later layers exhibit a significantly larger positive effect size, indicating that a higher cosine similarity in these layers strongly correlates with correct, grounded generation.

Leveraging these insights, we propose \textbf{Layer-Weighted Vector Steering (LWVS)}. Unlike uniform steering, LWVS applies an adaptive weight schedule that amplifies intervention in the influential later layers while reducing it in earlier layers. This design concentrates the steering effect where it is most mechanically relevant.

Our main contributions are summarized as follows:
\begin{itemize}
    \item We propose a novel \textbf{modality-aware vector steering} method that utilizes \textit{silent audio of the same length} to isolate acoustic information, distinguishing our approach from steering methods like VISTA \cite{vis} that use only text for negative instance.
    \item We design a probing framework using cosine similarity and Cohen's $d$ effect size to analyze internal model states. Our findings demonstrate that later layers have a disproportionately higher influence on output correctness.
    \item Based on this layer-wise analysis, we propose \textbf{Layer-Weighted Vector Steering (LWVS)}, a training-free intervention that adapts steering strength to influential layers, significantly mitigating hallucination while preserving general audio understanding capabilities.
\end{itemize}
\begin{figure}[t]
\includegraphics[width=0.45\textwidth]{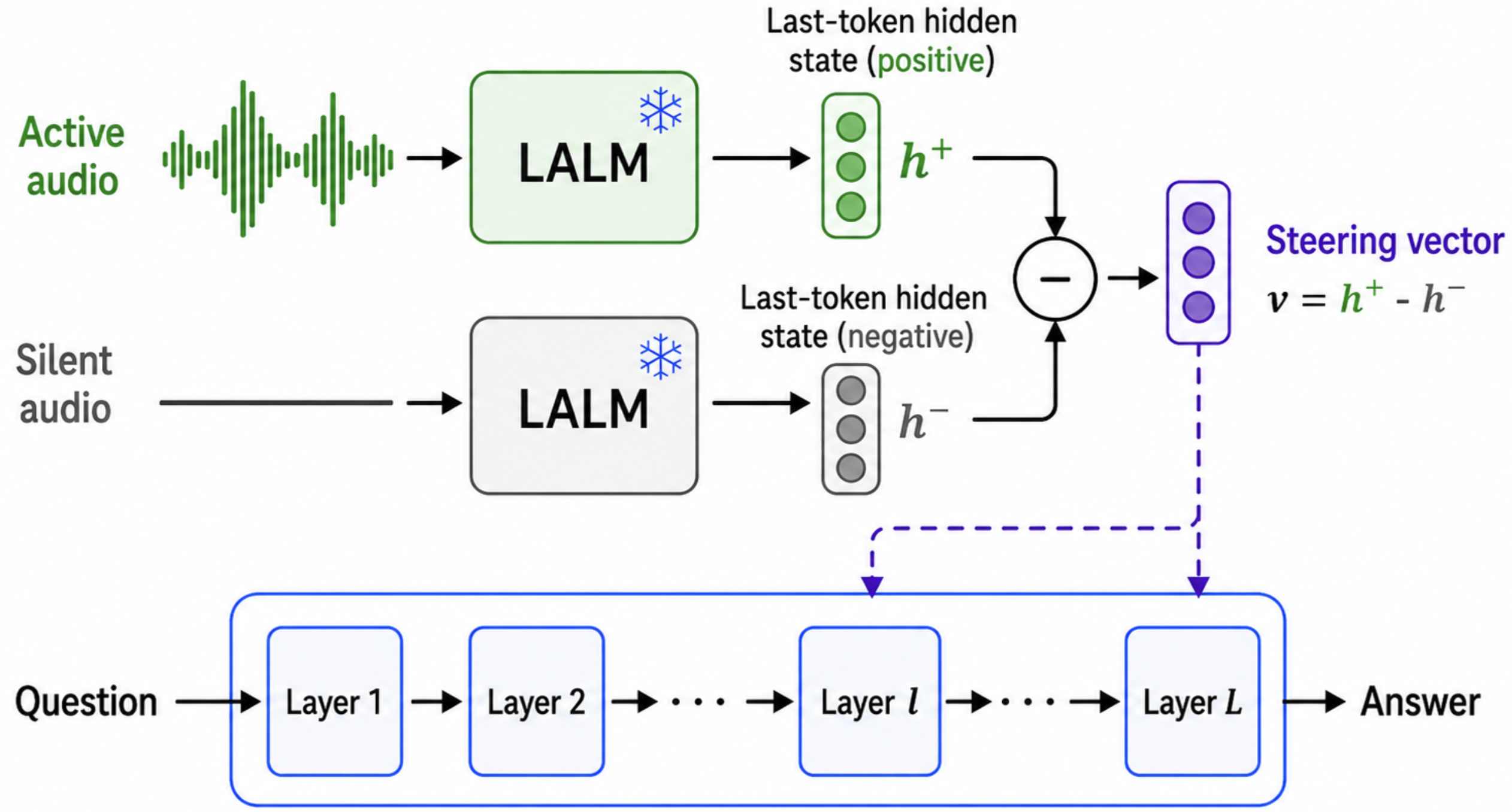}
\caption{\textbf{Modality-Aware Vector Steering} The steering vector is defined as the contrast between the last-token hidden states of the residual streams, computed by subtracting the representation of the negative instance(silent audio) from that of the positive instance.}
\label{fig:illustration}
\end{figure}

\section{Related Work}
Mitigating hallucinations in Large Language Models and Vision Langauge Models remains a critical challenge \cite{huang2025survey, li2023evaluating}. Conventional strategies, such as fine-tuning on domain-specific corpora or employing Retrieval-Augmented Generation (RAG) \cite{hu-etal-2025-removal}, often demand substantial data annotation or complex retrieval infrastructure. In contrast, our work explores inference-time intervention, offering a training-free and computationally efficient alternative.

\subsection{Activation Steering and VISTA}

Activation steering \cite{turner2023activation, zou2023representation} involves modifying a model's internal states during inference to guide its behavior. This technique has been successfully applied to LLMs to control stylistic attributes \cite{konen2024style} or to enhance truthfulness by identifying ``truth directions'' in activation space \cite{parksteer}.

Recently, VISTA \cite{vis} adapted this concept to Vision-Language Models (VLMs). VISTA constructs a steering vector by contrasting the activations of an image-text pair against a negative instance lacking image tokens. By adding this vector, the model is steered to prioritize visual context over language priors. However, VISTA typically applies a uniform steering strength across all layers. We differentiate our work by (1) extending this to the audio domain via a novel silence-based contrast, and (2) showing uniform steering is suboptimal, as hallucination influence is layer-dependent.

\subsection{Hallucination in Large Audio Models}
As Large Audio-Language Models (LALMs) proliferate, addressing their tendency to hallucinate has become a priority. Recent studies have characterized this phenomenon: \cite{nishimura2024audio} investigated audio hallucinations in Video-LLAMA, while \cite{kuan2024understanding} categorized object hallucinations in LALMs.

Current mitigations generally rely on decoding-time interventions \cite{li2023contrastive}. For instance, Audio-Aware Decoding (AAD) \cite{hsu2025reducing} employs contrastive decoding to penalize ungrounded tokens. While effective, AAD is computationally expensive, requiring two forward passes per generated token. In contrast, our proposed Layer-Weighted Vector Steering (LWVS) computes the steering vector once and adds it directly to the hidden states, incurring negligible latency while effectively mitigating hallucination.

\section{Methodology}
\label{sec:method}
In this section, we detail our approach to mitigating hallucination in Large Audio-Language Models. We begin in Section~\ref{ssec:VectorSteering} by addressing the limitations of directly applying standard text-based steering to audio. To overcome these shortcomings, we introduce a \textbf{Modality-Aware Steering Vector (MAVS)}, constructed by explicitly contrasting audio activations against a silence audio with equal duration. Next, to determine the optimal application of this vector, we probe the model's internal states in Section~\ref{ssec:Layer-wise}, identifying a significant layer-dependent correlation between vector similarity and output correctness. Finally, leveraging these empirical insights, we propose \textbf{Layer-Weighted Vector Steering (LWVS)} in Section~\ref{ssec:inference}, a targeted intervention that adaptively scales steering strength to maximize grounding.
\subsection{Internal state analysis}
\label{ssec:Layer-wise}
To determine the optimal application of our steering vector, we investigate how different layers contribute to the model's ability to distinguish between grounded and ungrounded content. We hypothesize that not all layers are equally sensitive to the acoustic evidence captured by our \textbf{modality-aware vector (defined in Section~\ref{ssec:VectorSteering})}.

\begin{figure*}[t!]
    \centering
    \includegraphics[width=0.95\textwidth]{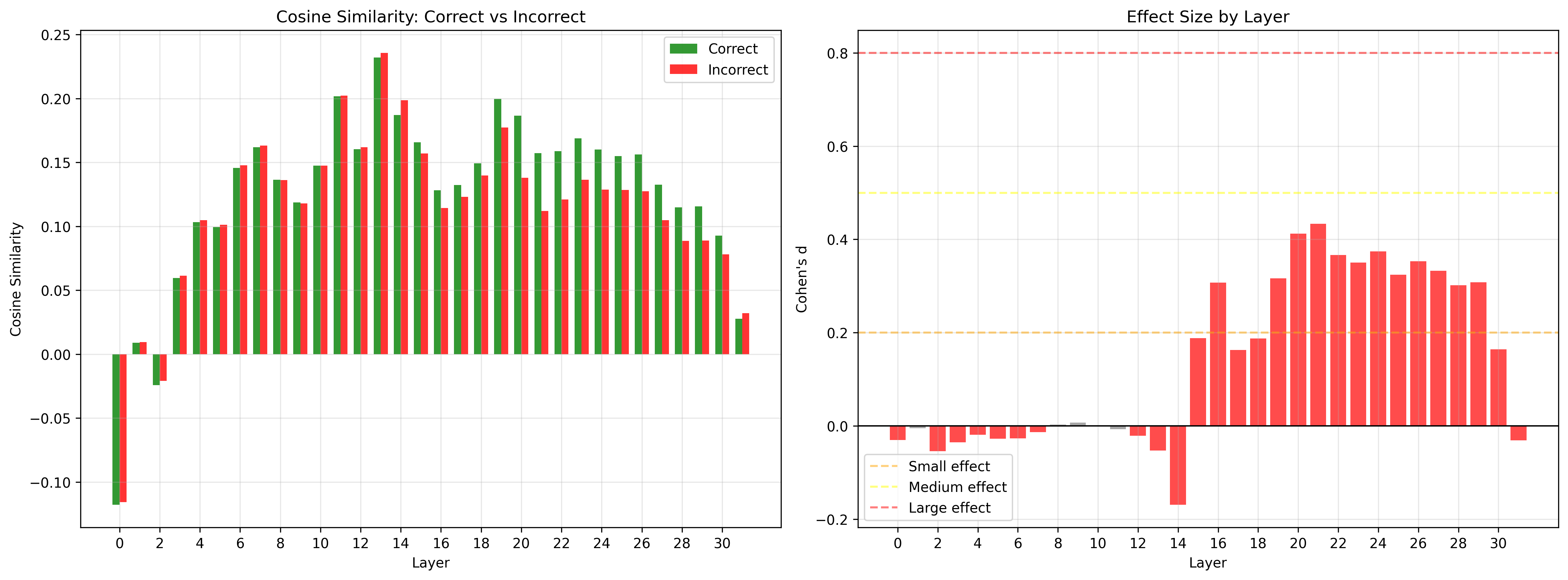}
    
    
    \includegraphics[width=0.95\textwidth]{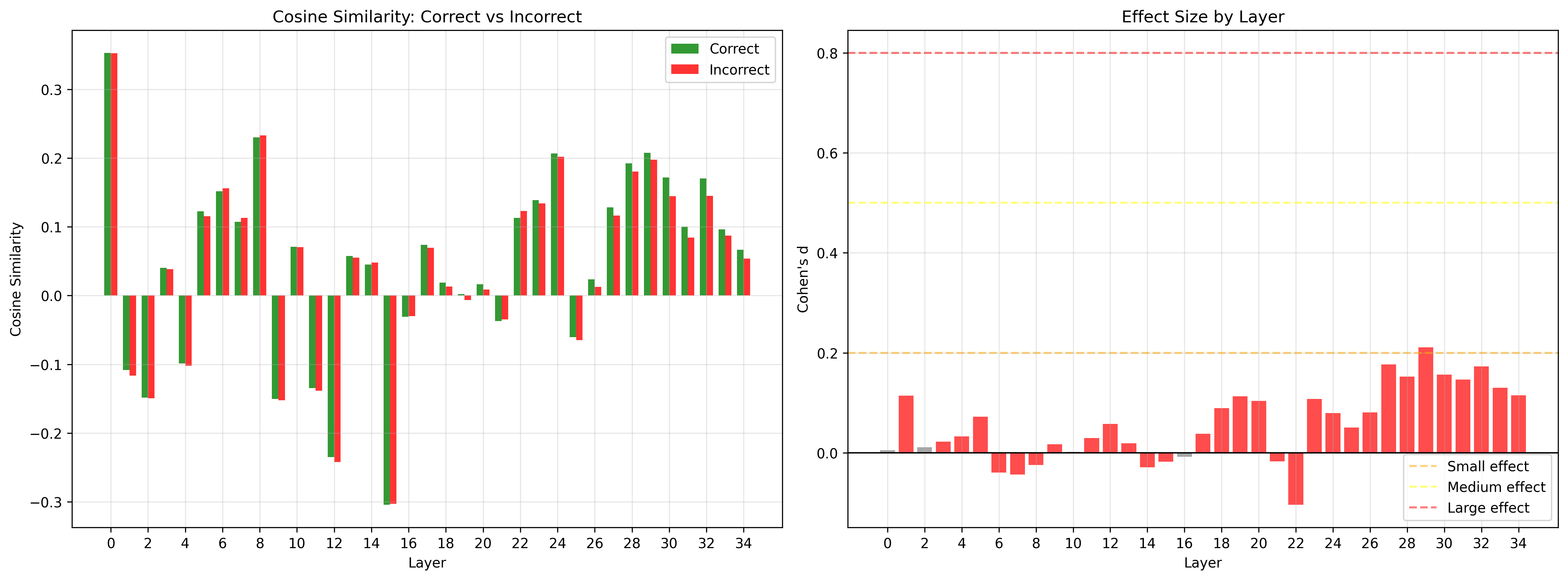}
    \caption{Layer-wise analysis for Gemma and Qwen. Top-Left: Cosine similarity of Qwen. Top-Right: Effect size of Qwen. Bottom-Left: Cosine similarity of Gemma. Bottom-Right: Effect size of Gemma.}
    \label{fig:Gemma-layers}
\end{figure*}

To quantify this, we categorize model generations into two groups: \textit{correct} (grounded) and \textit{incorrect} (hallucinated). For each group, we compute the cosine similarity between the model's internal hidden states $h_l$ at layer $l$ and the steering vector $v$. As shown in the left panels of Figure~\ref{fig:Gemma-layers}, we observe the distribution of these similarity scores across layers.

To rigorously measure the separation between these two distributions, we calculate Cohen's $d$ effect size for each layer (right panels). A higher effect size indicates a stronger distinction in the internal representation between grounded and ungrounded states. Our results reveal a consistent trend across both Qwen and Gemma models: while early layers show negligible separation, the effect size increases dramatically in the later layers.

This finding suggests that the model's critical decision-making regarding audio grounding occurs predominantly in the deeper layers. These empirical insights directly motivate our Layer-Weighted Vector Steering (LWVS) approach, which we detail in Section~\ref{ssec:inference}. Instead of uniform application, interventions should be concentrated in these high-influence layers to maximize efficacy.

\subsection{Modality-Aware Vector Steering}
\label{ssec:VectorSteering}

Standard steering approaches often rely on text-only contrasts, which fail to capture the acoustic nature of audio hallucinations. To address this, we introduce a \textbf{Modality-Aware Steering Vector} that explicitly isolates acoustic information.

We define a positive instance $X^+ = (x_a, x_q)$, consisting of the input audio $x_a$ and the user query $x_q$. Correspondingly, we construct a negative instance $X^- = (x_s, x_q)$, where $x_s$ is a \textit{silent} audio clip of the same duration as $x_a$. Let $F_l(X)$ denote the activation of the residual stream at the final token index $T$ for layer $l$. The steering vector $v^l$ is computed as the difference between the activations of the audio-grounded and silence-grounded instances:
\begin{equation}
v^l = F_l(X^+) - F_l(X^-), \quad l \in \{0, \dots, L-1\}.
\end{equation}
During inference, we inject this vector into the residual stream $h_{t,l}$ at every generation step $t$. To prevent the intervention from altering the magnitude of the model's features, we apply a normalization constraint:
\begin{equation}
\tilde{h}_{t,l} = \text{Norm}(h_{t,l} + \lambda^l v^l),
\end{equation}
where $\text{Norm}(u) = u \cdot \frac{\lVert h_{t,l} \rVert_2}{\lVert u \rVert_2}$ ensures that the norm of the modified hidden state matches the original.

\begin{table*}[t]
  \caption{Comparison of 4 methods using Gemma and Qwen. \textbf{Acc}:Accuracy, \textbf{Prec}:Precision, \textbf{Rec}:Recall. \textbf{TVS}: Text-Only Vector Steering, \textbf{MAVS}: Modality-Aware Vector Steering, \textbf{LWVS}:Layer-Weighted Vector Steering.}
  \label{tab:wide_results}
  \centering
  
  \scriptsize 
  \setlength{\tabcolsep}{2.5pt} 
  
  \begin{tabular*}{\textwidth}{@{\extracolsep{\fill}}cccccccccccccccccc}
    \toprule
    \multirow{2}{*}[-3pt]{\textbf{Method}} & \multirow{2}{*}[-3pt]{\textbf{Model}} & \multicolumn{4}{c}{\textbf{Random}} & \multicolumn{4}{c}{\textbf{Popular}} & \multicolumn{4}{c}{\textbf{Adversarial}} & \multicolumn{4}{c}{\textbf{Total}} \\
    \cmidrule(lr){3-6} \cmidrule(lr){7-10} \cmidrule(lr){11-14} \cmidrule(lr){15-18}
     & & Acc & Prec & Rec & F1 & Acc & Prec & Rec & F1 & Acc & Prec & Rec & F1 & Acc & Prec & Rec & F1 \\
    \midrule

    \multirow{2}{*}{\textbf{Original}} 
      & Gemma & 69.4 & 70.3 & 67.2 & 68.7 & 48.6 & 50.4 & 46.7 & 48.5 & 47.6 & 49.8 & 47.2 & 48.5 & 55.0 & 56.6 & 53.4 & 55.0 \\
      & Qwen  & 68.6 & 63.0 & 89.9 & 74.1 & 51.5 & 53.1 & 55.2 & 54.1 & 45.2 & 47.8 & 55.3 & 51.3 & 55.0 & 55.1 & 66.2 & 60.2 \\
    \addlinespace[4pt] 

    \multirow{2}{*}{\textbf{TVS}} 
      & Gemma & 68.5 & \textbf{75.6} & 54.6 & 63.4 & 47.2 & 48.7 & 34.1 & 40.1 & 47.5 & 49.6 & 34.6 & 40.8 & 54.3 & \textbf{57.7} & 40.8 & 47.8 \\
      & Qwen  & 70.8 & 64.9 & 90.6 & 75.7 & 50.1 & 51.8 & \textbf{55.6} & 53.6 & 44.6 & 47.4 & \textbf{55.7} & 51.2 & 55.0 & 55.1 & 66.8 & 60.4 \\
    \addlinespace[4pt]

    \multirow{2}{*}{\textbf{MAVS}(ours)} 
      & Gemma & \textbf{69.5} & 66.7 & 78.1 & 71.9 & 51.0 & 52.5 & 58.2 & 55.2 & 48.5 & 50.6 & 58.6 & 54.3 & 56.2 & 56.5 & 64.6 & 60.3 \\
      & Qwen  & 75.7 & 69.2 & 93.0 & 79.3 & 54.1 & 55.9 & 54.6 & 55.2 & 47.5 & 49.7 & 54.8 & 52.2 & 59.0 & 58.8 & 66.9 & 62.6 \\
    \addlinespace[4pt]

    \multirow{2}{*}{\textbf{LWVS}(ours)} 
      & Gemma & 69.3 & 65.4 & \textbf{81.9} & \textbf{72.7} & \textbf{51.5} & \textbf{52.7} & \textbf{62.8} & \textbf{57.3} & \textbf{48.9} & \textbf{50.8} & \textbf{63.1} & \textbf{56.3} & \textbf{56.4} & 56.2 & \textbf{69.0} & \textbf{61.9} \\
      & Qwen  & \textbf{77.3} & \textbf{70.6} & \textbf{93.7} & \textbf{80.5} & \textbf{54.7} & \textbf{56.6} & 54.6 & \textbf{55.5} & \textbf{48.1} & \textbf{50.3} & 54.8 & \textbf{52.4} & \textbf{59.9} & \textbf{59.7} & \textbf{67.1} & \textbf{63.2} \\
    \bottomrule
  \end{tabular*}
\end{table*}

\begin{table}[t]
  \centering
  \caption{Accuracy comparison on MMAU benchmarks. \textbf{TVS}: Text-Only Vector Steering, \textbf{MAVS}: Modality-Aware Vector Steering, \textbf{LWVS}: Layer-Weighted Vector Steering.}
  \label{tab:mmau_results}
  
  \setlength{\tabcolsep}{7pt}
  
  \begin{tabular}{ccccc}
    \toprule
    \multirow{2}{*}[-3pt]{\textbf{Method}} & \multicolumn{2}{c}{\textbf{MMAU Test}} & \multicolumn{2}{c}{\textbf{MMAU Test-mini}} \\
    \cmidrule(lr){2-3} \cmidrule(lr){4-5}
     & \textbf{Gemma} & \textbf{Qwen} & \textbf{Gemma} & \textbf{Qwen} \\
    \midrule
    
    \textbf{Original} & 63.8 & 54.8 & 66.1 & 56.4 \\
    \textbf{TVS}      & 63.1    & 58.3    & 65.5    & 59.2    \\
    \textbf{MAVS}(ours)      & \textbf{64.2} & 58.4 & \textbf{66.6} & 60.6 \\
    \textbf{LWVS}(ours)     & 64.1 & \textbf{59.2} & 66.4 & \textbf{61.4} \\
    \bottomrule
  \end{tabular}
\end{table}
\subsection{Layer-Weighted Vector Steering (LWVS)}
\label{ssec:inference}

Standard Vector Steering applies a uniform intervention strength ($\lambda^l = \lambda$) across all layers. However, our internal state analysis (Section~\ref{ssec:Layer-wise}) reveals that early layers contribute minimally to grounding, while later layers are highly influential. Furthermore, prior work suggests that lower layers in speech models encode speaker characteristics, whereas upper layers process phonetic and semantic content \cite{chung2019unsupervised}.

Based on these insights, we propose \textbf{Layer-Weighted Vector Steering (LWVS)}. We partition the model layers into a ``boost'' set $\mathcal{L}_{inc}$ (later layers) and a ``suppress'' set $\mathcal{L}_{dec}$ (early layers and the final output layers). We define a layer-specific strength schedule $\lambda^l$:
\begin{equation}
\lambda^l = 
\begin{cases} 
\lambda \cdot (1 + \frac{|\mathcal{L}_{dec}|}{|\mathcal{L}_{inc}|}\beta) & \text{if } l \in \mathcal{L}_{inc} \\
\lambda \cdot (1 - \beta) & \text{if } l \in \mathcal{L}_{dec}
\end{cases}
\end{equation}
where $\beta \in [0, 1]$ is a hyperparameter controlling the degree of adaptation. This design satisfies two critical constraints:
\begin{enumerate}
    \item \textbf{Energy Conservation:} The total steering effort remains constant ($\sum \lambda^l = L\lambda$), ensuring a fair comparison to uniform steering.
    \item \textbf{Logit Protection \& Empirical Alignment:} We exclude the final layers from the boost set to preserve the output distribution. Empirically, Figure \ref{fig:Gemma-layers} show that intervention efficacy diminishes or even becomes negative in these final stages compared to the peak intermediate layers.
\end{enumerate}

\section{Experimental Setup}
\subsection{Models}
We evaluate our method on two distinct multimodal architectures: \textbf{Qwen2-Audio-7B-Instruct} \cite{chu2024qwen2}, a specialized Large Audio-Language Model (LALM), and \textbf{Gemma-3n-E4B-It} \cite{team2025gemma}, a general Multimodal Large Language Model (MLLM). These models represent different architectural paradigms, allowing us to assess the generalizability of LWVS. Notably, Gemma utilizes the AltUp (Alternating Updates) transformer architecture. For this model, we apply steering vectors exclusively to the primary residual stream, leaving the auxiliary prediction and correction branches unmodified.

\subsection{Implementation Details}
We configure the layer-weighted hyperparameters based on the analysis in Section \ref{sec:method}.
\begin{itemize}
    \item \textbf{Qwen:} We define the boost set $\mathcal{L}_{inc} = \{14, \dots, 29\}$.
    \item \textbf{Gemma:} We define $\mathcal{L}_{inc} = \{16, \dots, 32\}$.
\end{itemize}
For both models, the remaining layers (early layers and the final two layers) form the suppression set $\mathcal{L}_{dec}$. We fix the base steering strength $\lambda = 0.05$ for all experiments. For our proposed \textbf{LWVS}, we set the adaptation factor $\beta = 0.5$, balancing the redistribution of steering energy towards the influential layers.
\subsection{Benchmarks}
We employ two benchmarks to evaluate hallucination mitigation and general capability preservation. Given that both tasks require discriminative answers (binary classification or multiple-choice selection), we utilize greedy decoding to strictly select the most probable answer token.

To measure object hallucination, we utilize the Audio Hallucination QA dataset, where the model must identify the presence of specific audio events. We prepend the instruction \textit{``Focus on the given audio and answer the following question''} and append \textit{``Answer with only yes or no.''}

To verify that our steering does not degrade general audio understanding, we evaluate on the Multi-Modal Audio Understanding (MMAU) benchmark\cite{sakshi2024mmau}. This suite consists of 10,000 expert-level multiple-choice questions. We append the instruction \textit{``Answer with ONLY the option letter ''} to constrain the generation to a single character.

\section{Results and Analysis}

In this section, we evaluate the effectiveness of our proposed methods in mitigating hallucination and preserving general audio understanding. We compare our Modality-Aware Vector Steering (MAVS) and Layer-Weighted Vector Steering (LWVS) against two baselines: the Original model (no steering) and Text-Only Vector Steering (TVS), which adapts standard VISTA-like text steering to audio.

\subsection{Audio Hallucination Mitigation}
Table~\ref{tab:wide_results} presents the performance on the Audio Hallucination QA dataset. The results highlight three key findings:
\begin{enumerate}
    \item \textbf{Text-centric steering is insufficient for audio.}
    Directly applying text-based steering (TVS) yields inconsistent results. While it offers marginal gains on Qwen, it significantly degrades performance on Gemma, dropping the Total F1 score from 55.0 to 47.8. This confirms our hypothesis that steering vectors must explicitly capture the acoustic modality to be effective in LALMs.
    \item \textbf{Modality-Aware steering provides robust grounding.}
    Our MAVS method, which constructs vectors via audio-silence contrast, consistently outperforms both the Original baseline and TVS. On Gemma, MAVS recovers the performance lost by TVS, achieving a Total F1 of 60.3 (+5.3 over baseline). On Qwen, it improves the Total F1 to 62.6. This demonstrates that anchoring the steering vector in the acoustic domain is essential for reducing ungrounded content.
    \item \textbf{Layer-Weighted Vector Steering (LWVS) achieves optimal performance.}
    By concentrating intervention on the high-influence layers identified in Section \ref{sec:method}, LWVS further refines the steering effect. It achieves the highest scores across the board, reaching a Total F1 of \textbf{61.9} on Gemma and \textbf{63.2} on Qwen. Notably, LWVS shows substantial improvements in the ``Random'' and ``Popular'' subsets, indicating that targeted steering effectively suppresses hallucinations even in challenging, diverse scenarios.
\end{enumerate}

\subsection{Impact on General Capabilities}
A critical concern with hidden-state intervention is the potential degradation of general reasoning capabilities. To assess this, we evaluate our methods on the MMAU benchmark, as shown in Table~\ref{tab:mmau_results}.

\noindent\textbf{Robustness across architectures.} 
For Gemma, LWVS maintains performance parity with the original model (64.1\% vs. 63.8\%), confirming that our targeted intervention does not disrupt the model's core multimodal reasoning.

\noindent\textbf{Synergy in Qwen.}
For Qwen, LWVS yields a significant performance boost, increasing MMAU accuracy from 54.8\% to \textbf{59.2\%}. This suggests that for some architectures, reducing hallucination has a positive downstream effect on general reasoning, likely by filtering out noise that would otherwise interfere with complex audio comprehension tasks.

\section{Conclusion}

In this work, we address the limitations of traditional text-only steering, which often fails to mitigate the unique hallucination patterns found in Large Audio-Language Models (LALMs). To overcome this, we propose modality-aware steering, a contrastive approach that constructs a steering vector $v$ by anchoring audio-derived internal states against a silence counterpart. Through extensive probing of the model’s internal states, we identified that the critical distinction between grounded and ungrounded content—measured via cosine similarity and Cohen's $d$ effect sizes—emerges predominantly within the deeper layers of the transformer stack. Building on this discovery, we introduce Layer-Weighted Vector Steering (LWVS), a training-free intervention that adaptively applies the modality-aware vector to these high-influence layers. Our evaluations on the Audio Hallucination QA and MMAU benchmarks demonstrate that LWVS significantly reduces ungrounded generation and preserves multimodal reasoning performance. Collectively, these findings provide a computationally efficient and interpretable framework for enhancing the reliability of audio-language models without the need for resource-intensive fine-tuning.

\section{Acknowledgment}
We gratefully acknowledge the National Center for High-performance Computing (NCHC) of the National Institutes of Applied Research (NIAR) in Taiwan for providing computational and storage resources.
This work was supported by the Ministry of Education (MOE) of Taiwan under the project Taiwan Centers of Excellence in Artificial Intelligence, through the NTU Artificial Intelligence Center of Research Excellence (NTU AI-CoRE).
We would like to thank Jen-Hao Cheng, Dau-Cheng Lyu, Tsung-Ying Yang, Hsiao-Tsung Hung, Chung-Cheng Chen and Chi-Yuan Hsiao at ASUS-OCIS for their helpful discussions and feedback throughout this work.

\section{Use of Generative AI Disclosure}
The authors disclose the use of generative AI tools (specifically, large language models) during the preparation of this manuscript. In accordance with ISCA policy, these tools were utilized solely for editing, polishing, and formatting text (such as condensing the abstract to meet submission length constraints) to improve readability. Generative AI was not used to produce any significant part of the scientific content, claims, or experimental results presented in this work. All co-authors remain fully responsible and accountable for the complete work and content of this paper and have consented to its submission.

\bibliographystyle{IEEEtran}
\bibliography{mybib}

@inproceedings{konen2024style,
  title={Style Vectors for Steering Generative Large Language Models},
  author={Konen, Kai and Jentzsch, Sophie Freya and Diallo, Diaoul{\'e} and Sch{\"u}tt, Peer and Bensch, Oliver and El Baff, Roxanne and Opitz, Dominik and Hecking, Tobias},
  booktitle={18th Conference of the European Chapter of the Association for Computational Linguistics, EACL 2024-Findings of EACL 2024},
  year={2024}
}

@article{goel2025audio,
  title={Audio flamingo 3: Advancing audio intelligence with fully open large audio language models},
  author={Goel, Arushi and Ghosh, Sreyan and Kim, Jaehyeon and Kumar, Sonal and Kong, Zhifeng and Lee, Sang-gil and Yang, Chao-Han Huck and Duraiswami, Ramani and Manocha, Dinesh and Valle, Rafael and others},
  journal={arXiv preprint arXiv:2507.08128},
  year={2025}
}

@article{team2025gemma,
  title={Gemma 3 technical report},
  author={Team, Gemma and Kamath, Aishwarya and Ferret, Johan and Pathak, Shreya and Vieillard, Nino and Merhej, Ramona and Perrin, Sarah and Matejovicova, Tatiana and Ram{\'e}, Alexandre and Rivi{\`e}re, Morgane and others},
  journal={arXiv preprint arXiv:2503.19786},
  year={2025}
}

@inproceedings{parksteer,
  title={Steer LLM Latents for Hallucination Detection},
  author={Park, Seongheon and Du, Xuefeng and Yeh, Min-Hsuan and Wang, Haobo and Li, Yixuan},
  booktitle={Forty-second International Conference on Machine Learning},
  year={2025}
}

@InProceedings{vis,
  title={The Hidden Life of Tokens: Reducing Hallucination of Large Vision-Language Models Via Visual Information Steering},
  author={Li, Zhuowei and Shi, Haizhou and Gao, Yunhe and Liu, Di and Wang, Zhenting and Chen, Yuxiao and Liu, Ting and Zhao, Long and Wang, Hao and Metaxas, Dimitris N},
  year = "2025",
  booktitle={Forty-second International Conference on Machine Learning}
}

@article{chu2024qwen2,
  title={Qwen2-Audio Technical Report},
  author={Chu, Yunfei and Xu, Jin and Yang, Qian and Wei, Haojie and Wei, Xipin and Guo, Zhifang and Leng, Yichong and Lv, Yuanjun and He, Jinzheng and Lin, Junyang and others},
  journal={CoRR},
  year={2024}
}

@article{nishimura2024audio,
  title={On the Audio Hallucinations in Large Audio-Video Language Models},
  author={Nishimura, Taichi and Nakada, Shota and Kondo, Masayoshi},
  journal={CoRR},
  year={2024}
}

@article{kuan2024understanding,
  title={Understanding Sounds, Missing the Questions: The Challenge of Object Hallucination in Large Audio-Language Models},
  author={Kuan, Chun-Yi and Huang, Wei-Ping and Lee, Hung-yi},
  journal={2024 Conference of the International Speech Communication Association (INTERSPEECH)},
  year={2024},
  arxiv = {2406.08402},
}

@article{chung2019unsupervised,
  title={An Unsupervised Autoregressive Model for Speech Representation Learning},
  author={Chung, Yu-An and Hsu, Wei-Ning and Tang, Hao and Glass, James},
  journal={Interspeech 2019},
  year={2019},
  publisher={ISCA}
}

@article{sakshi2024mmau,
  title={MMAU: A Massive Multi-Task Audio Understanding and Reasoning Benchmark},
  author={Sakshi, S and Tyagi, Utkarsh and Kumar, Sonal and Seth, Ashish and Selvakumar, Ramaneswaran and Nieto, Oriol and Duraiswami, Ramani and Ghosh, Sreyan and Manocha, Dinesh},
  journal={CoRR},
  year={2024}
}

@article{tseng2025taste,
  title={TASTE: Text-Aligned Speech Tokenization and Embedding for Spoken Language Modeling},
  author={Tseng, Liang-Hsuan and Chen, Yi-Chang and Lee, Kuan-Yi and Shiu, Da-Shan and Lee, Hung-yi},
  journal={arXiv preprint arXiv:2504.07053},
  year={2025}
}

@article{hsu2025reducing,
  title={Reducing Object Hallucination in Large Audio-Language Models via Audio-Aware Decoding},
  author={Hsu, Tzu-wen and Lu, Ke-Han and Chiang, Cheng-Han and Lee, Hung-yi},
  journal={arXiv preprint arXiv:2506.07233},
  year={2025}
}

@article{wu2025step,
  title={Step-Audio 2 Technical Report},
  author={Wu, Boyong and Yan, Chao and Hu, Chen and Yi, Cheng and Feng, Chengli and Tian, Fei and Shen, Feiyu and Yu, Gang and Zhang, Haoyang and Li, Jingbei and others},
  journal={arXiv preprint arXiv:2507.16632},
  year={2025}
}

@inproceedings{hu-etal-2025-removal,
    title = "Removal of Hallucination on Hallucination: Debate-Augmented {RAG}",
    author = "Hu, Wentao  and
      Zhang, Wengyu  and
      Jiang, Yiyang  and
      Zhang, Chen Jason  and
      Wei, Xiaoyong  and
      Qing, Li",
    editor = "Che, Wanxiang  and
      Nabende, Joyce  and
      Shutova, Ekaterina  and
      Pilehvar, Mohammad Taher",
    booktitle = "Proceedings of the 63rd Annual Meeting of the Association for Computational Linguistics (Volume 1: Long Papers)",
    month = jul,
    year = "2025",
    address = "Vienna, Austria",
    publisher = "Association for Computational Linguistics",
    url = "https://aclanthology.org/2025.acl-long.770/",
    doi = "10.18653/v1/2025.acl-long.770",
    pages = "15839--15853",
    ISBN = "979-8-89176-251-0"
}

@article{huang2025survey,
  title={A survey on hallucination in large language models: Principles, taxonomy, challenges, and open questions},
  author={Huang, Lei and Yu, Weijiang and Ma, Weitao and Zhong, Weihong and Feng, Zhangyin and Wang, Haotian and Chen, Qianglong and Peng, Weihua and Feng, Xiaocheng and Qin, Bing and others},
  journal={ACM Transactions on Information Systems},
  volume={43},
  number={2},
  pages={1--55},
  year={2025},
  publisher={ACM New York, NY}
}

@article{zou2023representation,
  title={Representation engineering: A top-down approach to {AI} transparency},
  author={Zou, Andy and Phan, Long and Chen, Sarah and Campbell, James and Guo, Phillip and Ren, Richard and Pan, Alexander and Yin, Xinyi and Mazeika, Mantas and Dombrowski, Ann-Kathrin and others},
  journal={arXiv preprint arXiv:2310.01405},
  year={2023}
}

@article{turner2023activation,
  title={Activation addition: Steering language models without optimization},
  author={Turner, Alex and Thiergart, Lisa and Udell, David and Leech, Gavin and Mini, Ulysse and MacDermott, Mac},
  journal={arXiv preprint arXiv:2308.10248},
  year={2023}
}

@inproceedings{li2023evaluating,
  title={Evaluating object hallucination in large vision-language models},
  author={Li, Yifan and Du, Yifan and Zhou, Kun and Wang, Jinpeng and Zhao, Wayne Xin and Wen, Ji-Rong},
  booktitle={Proceedings of the 2023 Conference on Empirical Methods in Natural Language Processing},
  pages={292--305},
  year={2023}
}

@inproceedings{li2023contrastive,
  title={Contrastive decoding: Open-ended text generation as optimization},
  author={Li, Xiang Lisa and Holtzman, Ari and Fried, Daniel and Liang, Percy and Zettlemoyer, Luke and Lewis, Mike},
  booktitle={Proceedings of the 61st Annual Meeting of the Association for Computational Linguistics (Volume 1: Long Papers)},
  pages={12286--12312},
  year={2023}
}

@inproceedings{pasad2021layer,
  title={Layer-wise analysis of a self-supervised speech representation model},
  author={Pasad, Ankita and Chou, Ju-Chieh and Livescu, Karen},
  booktitle={2021 {IEEE} Automatic Speech Recognition and Understanding Workshop ({ASRU})},
  pages={914--921},
  year={2021},
  organization={IEEE}
}

@article{chen2022wavlm,
  title={{WavLM}: Large-scale self-supervised pre-training for full stack speech processing},
  author={Chen, Sanyuan and Wang, Chengyi and Chen, Zhengyang and Wu, Yu and Liu, Shujie and Chen, Zhuo and Li, Jinyu and Kanda, Naoyuki and Yoshioka, Takuya and Xiao, Xiong and others},
  journal={{IEEE} Journal of Selected Topics in Signal Processing},
  volume={16},
  number={6},
  pages={1505--1518},
  year={2022}
}

@inproceedings{tang2024salmonn,
  title={{SALMONN}: Towards generic hearing abilities for large language models},
  author={Tang, Changli and Yu, Wenyi and Sun, Guangzhi and Chen, Xianzhao and Hu, Tianran and Lin, Chenghao and Zeng, Taesun and Liu, Shengpeng and Li, Chunping and Shen, Ting and others},
  booktitle={The Twelfth International Conference on Learning Representations},
  year={2024}
}

@inproceedings{deshmukh2023pengi,
  title={Pengi: An audio language model for audio tasks},
  author={Deshmukh, Soham and Elizalde, Benjamin and Singh, Raj and Wang, Huaming},
  booktitle={Advances in Neural Information Processing Systems},
  volume={36},
  pages={18092--18108},
  year={2023}
}

@article{defossez2024moshi,
  title={Moshi: a speech-text foundation model for real-time dialogue},
  author={D{\'e}fossez, Alexandre and Mazar{\'e}, Laurent and Orsini, Manu and Royer, Am{\'e}lie and P{\'e}rez, Patrick and J{\'e}gou, Herv{\'e} and Grave, Edouard and Zeghidour, Neil},
  journal={arXiv preprint arXiv:2410.00037},
  year={2024}
}

@article{zhang2023speechgpt,
  title={{SpeechGPT}: Empowering Large Language Models with Intrinsic Cross-Modal Conversational Abilities},
  author={Zhang, Dong and Li, Shimin and Zhang, Xin and Zhan, Jun and Wang, Pengyu and Zhou, Yaqian and Qiu, Xipeng},
  journal={arXiv preprint arXiv:2305.11000},
  year={2023}
}

@article{lakhotia2021generative,
  title={On generative spoken language modeling from raw audio},
  author={Lakhotia, Kushal and Kharitonov, Eugene and Hsu, Wei-Ning and Adi, Yossi and Polyak, Adam and Bolte, Benjamin and Nguyen, Tu-Anh and Copet, Jade and Baevski, Alexei and Mohamed, Abdelrahman and others},
  journal={Transactions of the Association for Computational Linguistics},
  volume={9},
  pages={1336--1354},
  year={2021}
}

@article{borsos2023audiolm,
  title={{AudioLM}: a Language Modeling Approach to Audio Generation},
  author={Borsos, Zal{\'a}n and Marinier, Rapha{\"e}l and Vincent, Damien and Kharitonov, Eugene and Pietquin, Olivier and Sharifi, Matt and Roblek, Dominik and Teboul, Olivier and Grangier, David and Tagliasacchi, Marco and others},
  journal={IEEE/ACM Transactions on Audio, Speech, and Language Processing},
  volume={31},
  pages={2523--2535},
  year={2023}
}

@article{wang2023neural,
  title={Neural codec language models are zero-shot text to speech synthesizers},
  author={Wang, Chengyi and Chen, Sanyuan and Wu, Yu and Zhang, Ziqiang and Zhou, Long and Liu, Shujie and Chen, Zhuo and Liu, Yanqing and Wang, Huaming and Li, Jinyu and others},
  journal={arXiv preprint arXiv:2301.02111},
  year={2023}
}

@article{fang2024llama,
  title={{LLaMA-Omni}: Seamless Speech Interaction with Large Language Models},
  author={Fang, Qingkai and Guo, Shoutao and Zhou, Yan and Ma, Zhengrui and Zhang, Shaolei and Feng, Yang},
  journal={arXiv preprint arXiv:2409.06666},
  year={2024}
}

\end{document}